\documentclass[a4paper]{jpconf}

\usepackage[english]{babel} 

\usepackage[%
	final,%
]{graphicx}

\usepackage[
   centertags, 
   sumlimits,  
   intlimits,  
   namelimits, 
]{amsmath} %
\usepackage{amssymb}

\usepackage[%
	pdftitle={Production, elliptic flow and energy loss of heavy quarks in the quark-gluon plasma},%
	pdfauthor={Jan Uphoff, Oliver Fochler, Zhe Xu, and Carsten Greiner},%
	pdfsubject={Production, elliptic flow and energy loss of heavy quarks at RHIC and LHC},
	pdfstartview=FitH,
	pdfpagemode=UseNone,
	bookmarksopen=true
	]{hyperref}

\usepackage[all]{hypcap} 

\usepackage{microtype} 

\makeatletter
\def\tagform@#1{\maketag@@@{\ignorespaces#1\unskip\@@italiccorr}}
\let\orgtheequation\theequation
\def\theequation{(\orgtheequation)}
\makeatother

\usepackage{xspace}

\newcommand{\beq}{\begin{equation}}
\newcommand{\eeq}{\end{equation}}

\newcommand{\jpsi}{$J/\psi$\xspace}
\newcommand{\vt}{$v_2$\xspace}
\newcommand{\RAA}{$R_{AA}$\xspace}

\begin{document}

\title{Open heavy flavor and J/psi at RHIC and LHC within a transport model}
\author{Jan Uphoff,$^1$ Oliver Fochler,$^1$ Zhe Xu,$^{2}$ and Carsten Greiner$^1$}

\address{$^1$ Institut f\"ur Theoretische Physik, Johann Wolfgang 
Goethe-Universit\"at Frankfurt, Max-von-Laue-Str. 1, 
D-60438 Frankfurt am Main, Germany}
\address{$^2$ Department of Physics, Tsinghua University, Beijing 100084, China}

\ead{uphoff@th.physik.uni-frankfurt.de}

\begin{abstract}
The production and space-time evolution of heavy quarks and $J/\psi$ in the quark gluon plasma is studied within the partonic transport model \emph{Boltzmann Approach to MultiParton Scatterings} (BAMPS). This framework allows interactions among all partons: gluons, light quarks and heavy quarks. Heavy quarks, in particular, interact with the rest of the medium via binary scatterings with a running coupling and a more precise Debye screening which is derived from hard thermal loop calculations. We compare our results of the elliptic flow and nuclear modification factor not only to experimental data of heavy flavor electrons at RHIC, but also to LHC data of heavy flavor electrons, muons, D mesons, and non-prompt $J/\psi$. Where no data is available yet, we make predictions for those observables. Furthermore, results on prompt $J/\psi$ elliptic flow are reported for RHIC energy within the same framework, taking the dissociation as well as regeneration of $J/\psi$ in the quark-gluon plasma into account.
\end{abstract}

\section{Introduction}

In ultra-relativistic heavy ion collisions at the Relativistic Heavy Ion Collider (RHIC) \cite{Adams:2005dq,Adcox:2004mh} and the Large Hadron Collider (LHC) \cite{Muller:2012zq} a unique state of matter is produced, in which quarks and gluons form the relevant degrees of freedom. This quark gluon plasma (QGP) has exciting properties such as collective behavior like a nearly perfect liquid or the quenching of high energy particles.

In particular heavy quarks (charm and bottom) provide an insightful way to learn more about the properties of this matter. Since they are heavy, their production time is at a very early stage of the heavy ion collision when enough energy is available \cite{Uphoff:2010sh}. In consequent interactions with other particles they are influenced by the medium without fully thermalizing \cite{Cao:2011et} and, therefore, carry information about its properties. Charm and bottom quarks travel through the medium, lose energy and participate in the collective motion. 
The exact mechanism of the intense interaction with the medium is actively debated, most recently, for instance, in \cite{Abir:2012pu,Meistrenko:2012ju,Uphoff:2012gb,He:2012df,Horowitz:2012cf,Lang:2012yf,Buzzatti:2012pe,Alberico:2012mv, Cao:2012au, Gossiaux:2012ya}.

Complimentary to these open heavy flavor particles are particles in which the flavor is hidden. That is, bound states of a heavy quark and an anti-heavy quark. They can survive in the QGP to some extent, but melt if the temperature of the medium is too high \cite{Mocsy:2007jz}. Therefore, they could be used as a thermometer. In addition to this melting they can also be regenerated by two heavy quarks which meet in the medium. The most prominent hidden heavy flavor particle is the \jpsi, which consists of a charm and an anti-charm quark.

In this paper we employ the partonic transport model BAMPS to study the suppression and elliptic flow of both open and hidden heavy flavor. In detail, we make calculations for open heavy flavor observables such as the nuclear modification factor \RAA and elliptic flow \vt of heavy flavor muons and electrons, D mesons and non-prompt \jpsi which stem from B mesons. Furthermore we present \vt calculations of prompt \jpsi.

\section{Partonic transport model BAMPS}
For the simulation of the QGP we use the partonic transport model \emph{Boltzmann Approach to MultiParton Scatterings} (BAMPS) \cite{Xu:2004mz,Xu:2007aa}, which describes the full space-time evolution of the QGP by solving the Boltzmann equation,
\begin{equation}
\label{boltzmann}
\left ( \frac{\partial}{\partial t} + \frac{{\mathbf p}_i}{E_i}
\frac{\partial}{\partial {\mathbf r}} \right )\,
f_i({\mathbf r}, {\mathbf p}_i, t) = {\cal C}_i^{2\rightarrow 2} + {\cal C}_i^{2\leftrightarrow 3}+ \ldots  \ ,
\end{equation}
for on-shell partons and pQCD interactions. On the heavy flavor sector all possible binary collisions are implemented, such as $
        g g  \leftrightarrow Q  \bar{Q} $, $
        q  \bar{q} \leftrightarrow Q  \bar{Q} $, $
        g Q \rightarrow g Q $, $
        q Q \rightarrow q Q$, and $
        J/\psi +g \leftrightarrow c + \bar c$. 
As a note, heavy quark production via gluon fusion within BAMPS is discussed in Ref.~\cite{Uphoff:2010sh}, but will not be addressed in this paper.
        
The divergent $t$ channel of elastic processes is regularized with a screening mass $\mu$ which is determined by matching energy loss calculations within the hard thermal loop approach \cite{Gossiaux:2008jv,Peshier:2008bg,Uphoff:2011ad}. Furthermore, the running of the coupling is explicitly taken into account.
We note that currently for light partons neither a running coupling nor an improved Debye screening is employed, which we plan, however, to do in the future.
Inelastic processes $2 \rightarrow 3$ are currently being studied for heavy quarks, but have already been implemented for light partons.
Details can be found in the following references: the model itself \cite{Xu:2004mz,Xu:2007aa}, the open heavy flavor implementation \cite{Uphoff:2010sh,Uphoff:2011ad,Uphoff:2012gb}, and the \jpsi details \cite{Uphoff:2011fu,Uphoff:2012it}.

\section{Results and comparison with data}

The implemented running coupling and the improved screening procedure, which reproduces the energy loss from hard thermal loop calculations, effectively enhances the heavy quark cross section with the medium. Quantitative comparisons \cite{Uphoff:2010bv,Uphoff:2011ad,Fochler:2011en,Meistrenko:2012ju,Uphoff:2012gb} show that elastic processes contribute significantly to the energy loss of heavy quarks. However, they alone can reproduce neither the data of the nuclear modification factor nor the elliptic flow of any heavy flavor particle species. This is not too surprising since we expect that radiative $2 \rightarrow 3$ processes also play an important role and that both processes together should account for the measured suppression and flow. The quantitative contribution of radiative processes will be studied in an forthcoming investigation. 

\begin{figure}[t]
\begin{minipage}[t]{0.49\textwidth}
\centering
\includegraphics[width=1.0\textwidth]{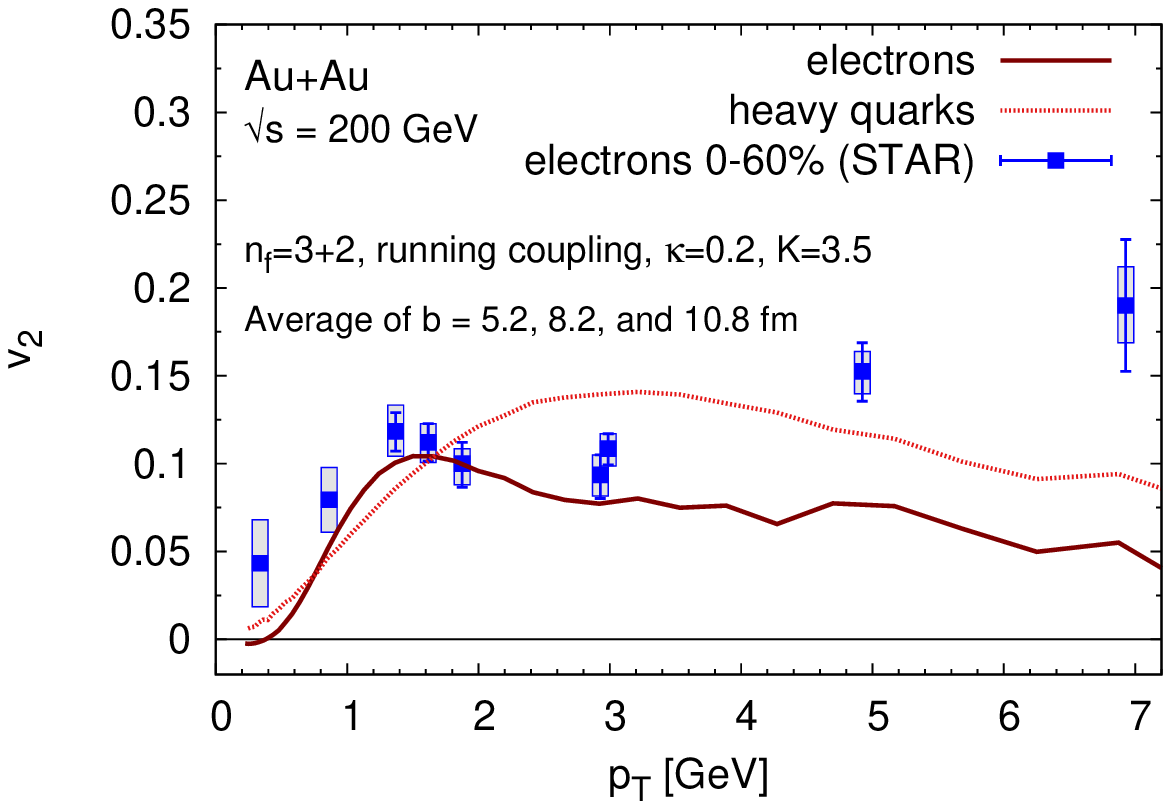}
\end{minipage}
\hfill
\begin{minipage}[t]{0.49\textwidth}
\centering
\includegraphics[width=1.0\textwidth]{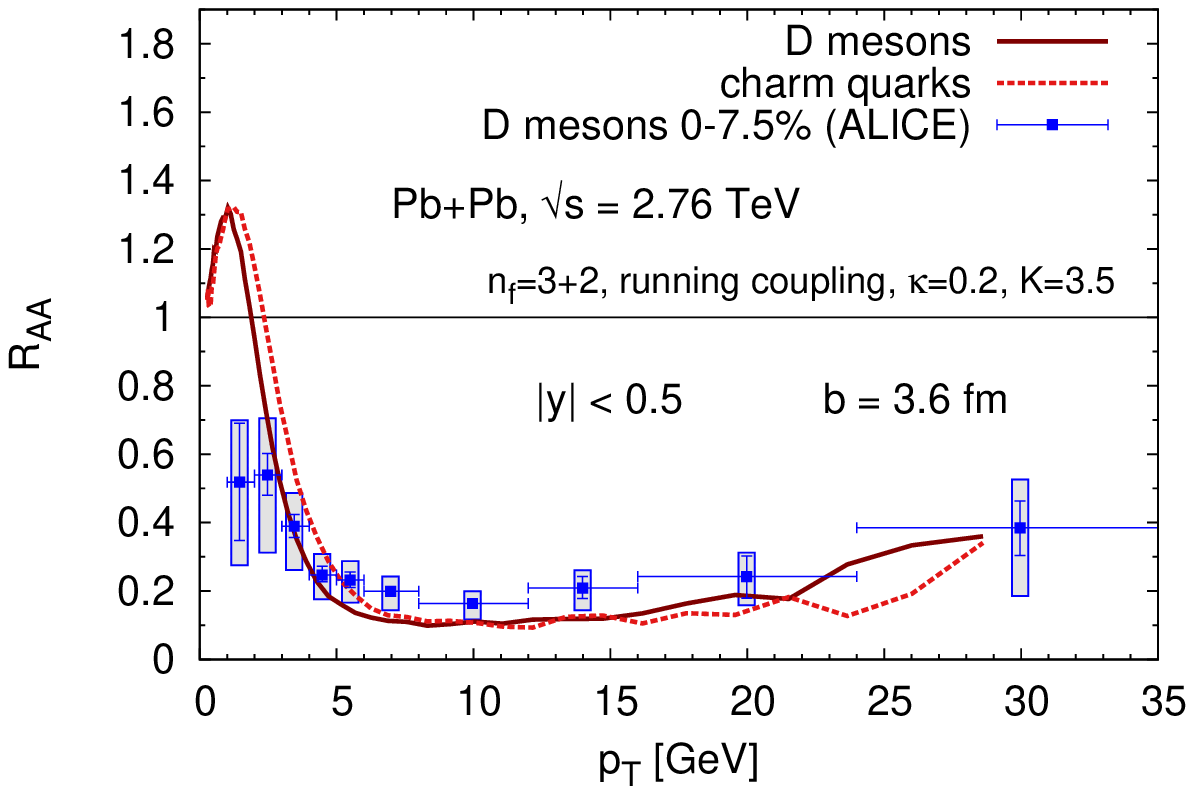}
\end{minipage}
\caption{Left: Elliptic flow $v_2$ of heavy flavor electrons at RHIC for 0-60\,\% centrality class  as a function of transverse momentum with STAR data \cite{Mustafa:2012jh}. Right: Nuclear modification factor $R_{AA}$ of $D$ mesons for $b=3.6\,{\rm fm}$ with data \cite{:2012yv}.
}
\label{fig:v2_rhic_raa_lhc}
\end{figure}

In this paper we mimic the radiative influence by effectively increasing the elastic cross section by a factor $K=3.5$ which fits the heavy flavor electron elliptic flow data from PHENIX in the centrality class 20-40\,\% \cite{Uphoff:2012gb}. 
On the left hand side of Fig.~\ref{fig:v2_rhic_raa_lhc} we show our prediction for 0-60\,\% centrality class for the same parameters, which is compared to new data from STAR \cite{Mustafa:2012jh}. 
Not surprisingly the agreement is very well for small and intermediate transverse momenta. Even the bump around $1.7\, {\rm GeV}$ is nicely described. At large $p_T$, however, the data is significantly larger than our calculations. This could also be due to jet-like correlations which are not considered in the analysis \cite{Mustafa:2012jh}. Simultaneously, the $R_{AA}$ data at RHIC can also be described with the same parameter \cite{Uphoff:2012gb}. Whether this effective description is valid and radiative contributions really boil down to simply multiplying the elastic cross section by a constant factor will be studied with BAMPS in the future.

On the right hand side of Fig.~\ref{fig:v2_rhic_raa_lhc} the nuclear modification factor of $D$ mesons for very central events at LHC is depicted. Our prediction with BAMPS is slightly smaller than the experimental data points. This is in agreement with the observation that we also underestimate the \RAA of non-prompt \jpsi, heavy flavor electrons, and muons \cite{Uphoff:2012gb}.

\begin{figure}[t]
\begin{minipage}[t]{0.49\textwidth}
\centering
\includegraphics[width=1.0\textwidth]{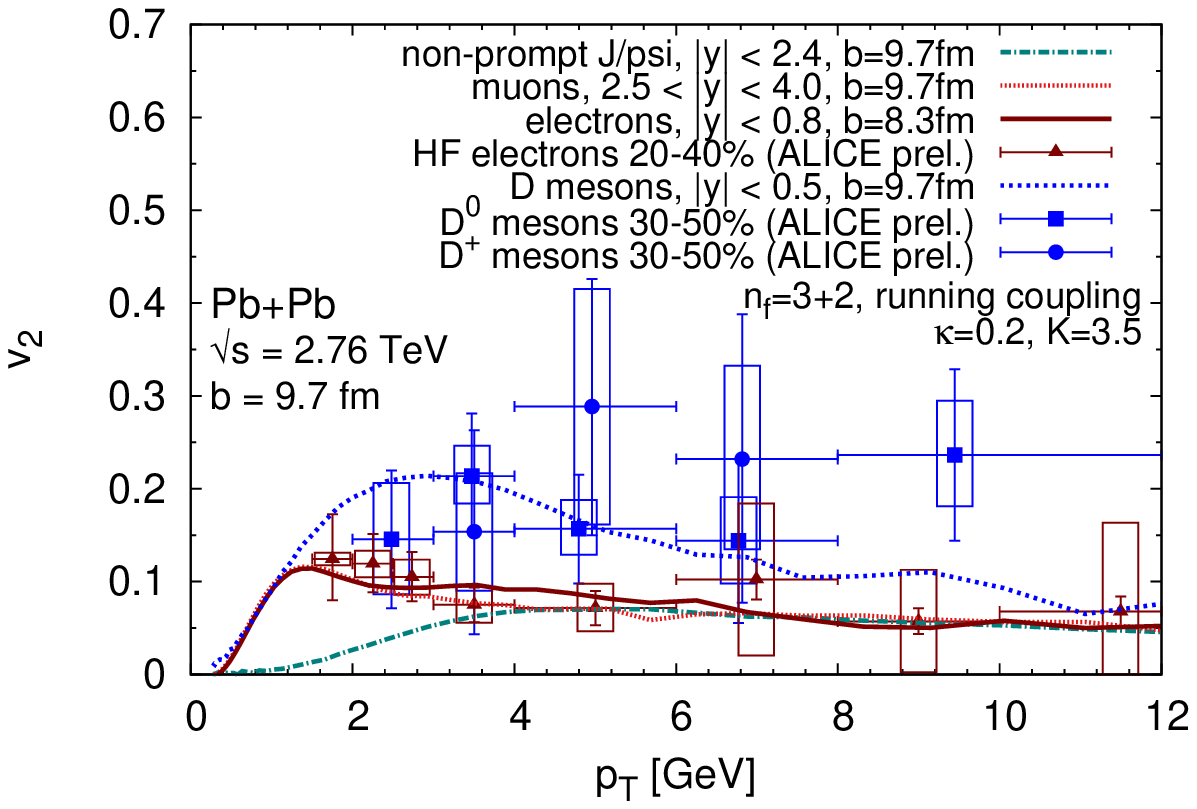}
\end{minipage}
\hfill
\begin{minipage}[t]{0.49\textwidth}
\centering
\includegraphics[width=1.0\textwidth]{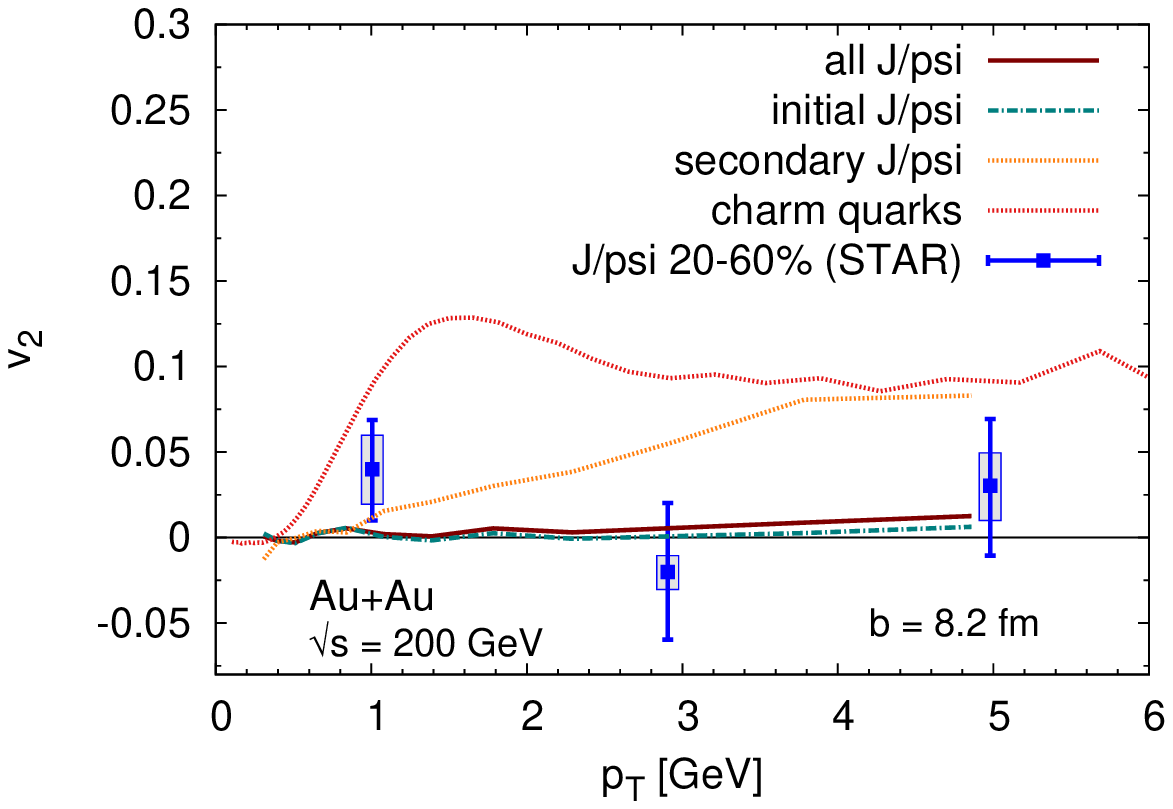}
\end{minipage}
\caption{Left: Elliptic flow $v_2$ of $D$ mesons, non-prompt $J/\psi$, muons, and electrons at LHC together with available data for $D$ mesons and electrons \cite{Ortona:2012gx,sakai:qm12}. Right: Elliptic flow of $J/\psi$ with data \cite{Powell:2011np}. For comparison the charm quark $v_2$ is also shown.
}
\label{fig:v2_lhc_v2_jpsi}
\end{figure}

The left hand side of Fig.~\ref{fig:v2_lhc_v2_jpsi} shows our predictions of the elliptic flow of $D$ mesons, non-prompt \jpsi, heavy flavor electrons, and muons. For $D$ mesons and electrons there is already data available which agrees very well with our calculations.

Experimental measurements at RHIC showed that the $v_2$ of prompt $J/\psi$ is very small \cite{Powell:2011np}. This is in contradiction with the regeneration picture where the flow of the charm quarks should be transferred to the $J/\psi$. BAMPS is an ideal framework to study this in more detail since it reproduces the open heavy flavor flow and also allows recombination of charm quarks to $J/\psi$.
The right panel of Fig.~\ref{fig:v2_lhc_v2_jpsi} shows that in BAMPS even with regeneration the elliptic flow of all $J/\psi$ is compatible with the 
data.

\section{Summary}

We presented heavy flavor calculations with the parton cascade BAMPS. So far only elastic interactions with a running coupling and an improved Debye screening are implemented on the heavy flavor sector. To account for missing radiative contributions we scale the elastic cross section with $K=3.5$ which is determined from comparing to the PHENIX \vt data at RHIC. With this parameter we find also a good agreement to recent \vt data from STAR in another centrality class. The predictions for all open heavy flavor \RAA at the LHC consistently underestimate the data slightly. However, the predicted \vt of D mesons and electrons is in very good agreement with recent data from ALICE. The radiative processes \cite{Abir:2011jb} are currently being implemented in BAMPS. It will be interesting to see if they can indeed account for scaling the binary interactions with a constant factor.
Furthermore, we showed some first BAMPS calculations on prompt \jpsi \vt at RHIC.

\section*{Acknowledgements}
J.U. is grateful for the kind hospitality at Tsinghua University, where part of this work has been done. Furthermore, J.U. would like to thank the DAAD and the Helmholtz Research School for Quark Matter studies for financial support. O.F. acknowledges support by the BMBF.

The BAMPS simulations were performed at the Center for Scientific Computing of the Goethe University Frankfurt. This work was supported by the Helmholtz International Center for FAIR within the framework of the LOEWE program launched by the State of Hesse.

\section*{References}
\bibliography{hq}

\end{document}